\documentclass[11pt]{article}
\usepackage{xspace}
\usepackage{graphicx}
\usepackage{amsmath}
\usepackage{fullpage}
\usepackage{amssymb}
\usepackage{color}
\usepackage{multirow,soul}
\def\Title#1{\begin{center} {\Large {\bf #1} } \end{center}}

\begin{document}

\Title{Mixing angle and phase predictions from $A_5$ with generalised CP}

\bigskip\bigskip


\begin{raggedright}  

{\it Jessica Turner\index{Turner. J},\\
Institute for Particle Physics Phenomenology, \\
Durham University,\\
South Road,  DH1 3LE Durham, UK}\\

\end{raggedright}
\vspace{1.cm}

{\small
\begin{flushleft}
\emph{To appear in the proceedings of the Prospects in Neutrino Physics Conference, 15 -- 17 December, 2014, held at Queen Mary University of London, UK.}
\end{flushleft}
}

\section{Introduction}
Models which use flavour symmetries have provided an appealing explanation of the pattern of mixing found in the lepton sector. In the following, fully presented in \cite{ballett}, we consider a non-Abelian discrete flavour  and  generalised CP (gCP)  as a symmetry at the high energy scale. At the  mass generation scale, this symmetry is broken into Abelian residual symmetries. From considerations of the residual symmetries a PMNS matrix can be constructed and  six observables (three mixing angles and three phases) of the lepton sector can be predicted. The flavour group considered is the alternating group on 5 elements, $A_5$. 

\section{Flavour symmetry}
We  assume  at the high energy scale a discrete, non-Abelian flavour symmetry, $G_{f}$, is present and the  three flavours are unified into three-dimensional irreducible representations of the flavour symmetry. As lepton masses are known to be distinct, the non-Abelian flavour symmetry cannot be a symmetry present at the low energy scale and thus the flavour group must be broken into Abelian residual symmetries present in the charged lepton, $G_e$, and neutrino, $G_\nu$, sectors.  Let $\rho\left(g_\nu\right)$ and $\rho\left(g_e\right)$ be three-dimensional representations of general elements of the Abelian residual symmetries $G_\nu$ and $G_e$ respectively. The left  handed charged lepton and neutrino transform according to
\begin{equation} \label{ressym} e_\text{L} \to \rho\left(g_e\right) e_\text{L}\quad\text{and}\quad\nu_\text{L}\to \rho\left(g_\nu\right) \nu_\text{L}.  \end{equation}
Invariance under the transformations of Eq.(\ref{ressym}) leads to the following constraints on the charged lepton and neutrino mass matrices 
\begin{equation}\label{massmatrix}m_\lambda m_\lambda^\dagger = \rho\left(g_e\right)^\dagger (m_\lambda
m_\lambda^\dagger)\rho\left(g_e\right) \quad\text{and}\quad m_\nu = \rho\left(g_\nu\right)^\text{T}m_\nu
\rho\left(g_\nu\right). \end{equation}
The  residual symmetries of  Eq.(\ref{massmatrix}) constrain the form of the mass matrices and therefore the structure of the PMNS matrix. The possible forms of residual symmetries can be derived from  two considerations: $G_{\nu}$ and $G_{e}$ must be Abelian subgroups of the non-Abelian flavour group, $G_{f}$, and the residual symmetries must be consistent with the largest symmetries allowed by  the mass terms.  In the diagonal basis, the largest symmetry of the charged lepton mass matrix is ${U(1)}^{3}$. Therefore a discrete residual symmetry in the charged lepton sector must be a direct product of cyclic groups, $G_{e}=Z_{m}$. As we have assumed neutrinos are Majorana in nature the largest residual symmetry possible is $\mathbb{Z}_2\times\mathbb{Z}_2$ and thus $G_{\nu}=\mathbb{Z}_2\times\mathbb{Z}_2$ or $G_{\nu}=\mathbb{Z}_2$.

\section{Generalised CP symmetry}
A gCP symmetry is a combination of charge conjugation and parity transformation which also acts on flavour indices.  The gCP transformation  on a set of fields, $\Psi$,
\[ \label{gCP}\Psi \to X\Psi^c,   \]
where $X$ is a unitary, symmetric matrix and $\Psi^c$ denotes the CP conjugate of $\Psi$. It was shown in \cite{ Holthausen:2012dk,Feruglio:2012cw} the  flavour and gCP symmetry must satisfy the following consistency equation
 \begin{equation} \label{consistency} X {\rho(g)}^{*}{X}^{*}=\rho(g^\prime),\end{equation}
where $g$ and $g^{\prime}$ are elements of $G_{f}$. It can be shown that this mapping  defines a group automorphism. Moreover in \cite{Chen}, they showed that  a  physical gCP transformation requires the additional property that gCP acts as a \emph{class inverting automorphism}. This implies that $g^{\prime}$ is conjugate to $g^{-1}$.  gCP invariance  places constraint on the mass matrices,
\[ X^\text{T} m_\nu X = m^*_\nu \quad\text{and}\quad X^\dagger
(m_\lambda  m^\dagger_\lambda) X = (m_\lambda
m^\dagger_\lambda)^*. \]
In the following,  we  assume gCP is broken in the charged lepton sector but remains unbroken in the neutrino sector.

\section{Constructing the PMNS matrix from symmetry constraints}
Both the flavour and gCP symmetries place constraints on the neutrino and charged lepton mass matrices.  From these constraints, the structure of the diagonalising matrices, $U_{\nu}$ and $U_{l}$,  of these mass matrices can be deduced and thus the  PMNS matrix can be constructed.  
 Consider the constraint on the charged lepton sector rephrased in terms of a commutator: $[\rho(g_{e}),(m_\lambda m^\dagger_\lambda)] = 0$. Using the unitarity of $\rho(g_{e}),$ and the hermiticity of $m_\lambda m^\dagger_\lambda$, there exists a basis in which $U_{l}$ simultaneously diagonalises residual symmetry $\rho(g_{e})$ and $m_\lambda m^\dagger_\lambda$,
\begin{equation} \label{Ue}U_{l}=U_{e}R_{e}(\theta,\gamma)\end{equation}
where $U_{e}$ is a diagonalising matrix of $\rho(g_{e})$ and $R_{e}(\theta,\gamma)$ is a complex rotation in the degenerate subspace of $\rho(g_{e})$, if such a subspace exists. 
In the neutrino sector, we consider the Abelian residual symmetry to be $\mathbb{Z}_2\times CP$. To determine the diagonalising matrix of the neutrino mass matrix we must consider the flavour residual symmetry, the gCP symmetry and the consistency between these symmetries. 
As $X$ is a symmetric, unitary matrix it can be decomposed as $X=\Omega^{\dagger}\Omega^{*}$, where $\Omega$ is unitary. Using $\Omega$, a convenient basis change can be performed such that the   neutrino mass matrix is real. The most general form of $U_{\nu}$ is
\begin{equation}\label{Unu} U_{\nu}=\Omega R_{\nu}(\theta)  \end{equation}
where $R_{\nu}(\theta)$ a rotation in the plane of the degenerate eigenspace of  $\rho(g_\nu)$. Combining Eq.(\ref{Ue}) and Eq.(\ref{Unu}) the PMNS matrix can be written as $U=R_{e}(\theta,\gamma){U_{e}}^{\dagger}\Omega R_{\nu}(\theta)$.
For an  in depth discussion of the derivation of $\Omega$ and also the group theoretic considerations see \cite{ballett}.  As we will consider the residual symmetries of the charged lepton sector to be   $\mathbb{Z}_5$, $\mathbb{Z}_3$, $\mathbb{Z}_2\times\mathbb{Z}_2$ there is no degenerate eigenspace and therefore no  additional complex rotation is needed and   the PMNS matrix simplifies to 
\[U=U_{e}^{\dagger}\Omega R_\nu(\theta).\] 
For each configuration of the PMNS matrix the arbitrariness of ordering of the diagonalising matrices must be accounted for and thus all rows and columns are permuted. The mixing angles  derived from these permuted matrices are then compared to global data \cite{Gonzalez-Garcia:2014bfa}.

\begin{table}[!t]
\begin{center}
\caption{\label{tab:predictions} Predictions of correlations as a function of dimensionless parameter $r \equiv
\sqrt{2}\sin\theta_{13}$.}
\resizebox{0.4\linewidth}{!}{
\begin{tabular}{ c | c|c|c|c  }  \hline\hline
$G_e$ & $\theta_{12} $ & $\theta_{23}$ & $\sin\alpha_{ji}$& $\delta$   \\
\hline
\multirow{2}{*}{$\mathbb{Z}_3$} & \multirow{2}{*}{$35.27^\circ + 10.13^\circ\, r^2$} & \multirow{2}{*}{$45^\circ$} &  \multirow{2}{*}{$0$}& $90^\circ$ \\
\cline{5-5}
 &  &  & & $270^\circ$ \\
\hline
\multirow{4}{*}{$\mathbb{Z}_5$} & \multirow{4}{*}{$31.72^\circ + 8.85^\circ\,r^2$} & \multirow{2}{*}{$45^\circ \pm 25.04^\circ\,r$} & \multirow{2}{*}{$0$} &$0^\circ$ \\
\cline{5-5}
 &  &  & & $180^\circ$ \\
\cline{3-5}
&  & \multirow{2}{*}{$45^\circ$} & \multirow{2}{*}{$0$} & $90^\circ$ \\
\cline{5-5}
 &  &  & & $270^\circ$  \\
\hline
\multirow{4}{*}{$\mathbb{Z}_2\times\mathbb{Z}_2$} & \multirow{4}{*}{$36.00^\circ - 34.78^\circ\,r^2$} & \multirow{2}{*}{$31.72^\circ + 55.76^\circ\,r$} & \multirow{4}{*}{$0$}&$0^\circ$ \\
\cline{5-5}
 &  &  & & $180^\circ$  \\
\cline{3-3}
\cline{5-5}
&  & \multirow{2}{*}{$58.28^\circ - 55.76^\circ\,r$} & & $0^\circ$ \\
\cline{5-5}
 &  &  & & $180^\circ$ \\  \hline\hline
\end{tabular}
}
\end{center}
\end{table}

\section{Results}
 Our predictions can be classified by the charged lepton residual symmetry $\mathbb{Z}_3$, $\mathbb{Z}_5$ and $\mathbb{Z}_2\times\mathbb{Z}_2$.  For each prediction the observables  are a function of the continuous parameter, $\theta$,  and so the mixing angles can be plotted as a function of $\theta$ as shown below in Figure~\ref{fig:Z5}. \\
In Table~\ref{tab:predictions} the continuous parameter, $\theta$, has been eliminated and the correlations between mixing angles can be found. The $\mathbb{Z}_3$ prediction has maximal $\theta_{23}$ and maximal $\delta$.  In addition, its $\theta_{12}$ value lies close to the upper region of the  3$\sigma$ global fit data. There are two distinct $\mathbb{Z}_5$ predictions: they share a common $\theta_{12}$ prediction which is close to the lower boundary of 3$\sigma$ data and they differ in that one predicts maximal $\theta_{23}$ and maximal $\delta$ whilst the other has non-maximal $\theta_{23}$ and CP conserving $\delta$. There are  two patterns produced by a preserved $\mathbb{Z}_2\times\mathbb{Z}_2$: the $\theta_{12}$ prediction they share is closer to the lower 3$\sigma$ and the non-maximal $\theta_{23}$ prediction is associated to CP conserving $\delta$. Moreover, all Majorana phases are  CP conserving. It is a striking signature that four of the eight predicts have maximal $\delta$ associated to maximal $\theta_{23}$.  Testing this pattern is an achievable goal by long base-line accelerator experiments such as T2K \cite{T2K} and NO$\nu$A \cite{NOVA}. In conjunction, the correlations between mixing angles, in particular correlations between $\theta_{12}$ and $\theta_{13}$, will be probed by medium-baseline reactor experiments  such as JUNO \cite{JUNO} and RENO-50 \cite{RENO} .

\begin{figure*}[!t]
\centering
\includegraphics[width=0.4\textwidth]{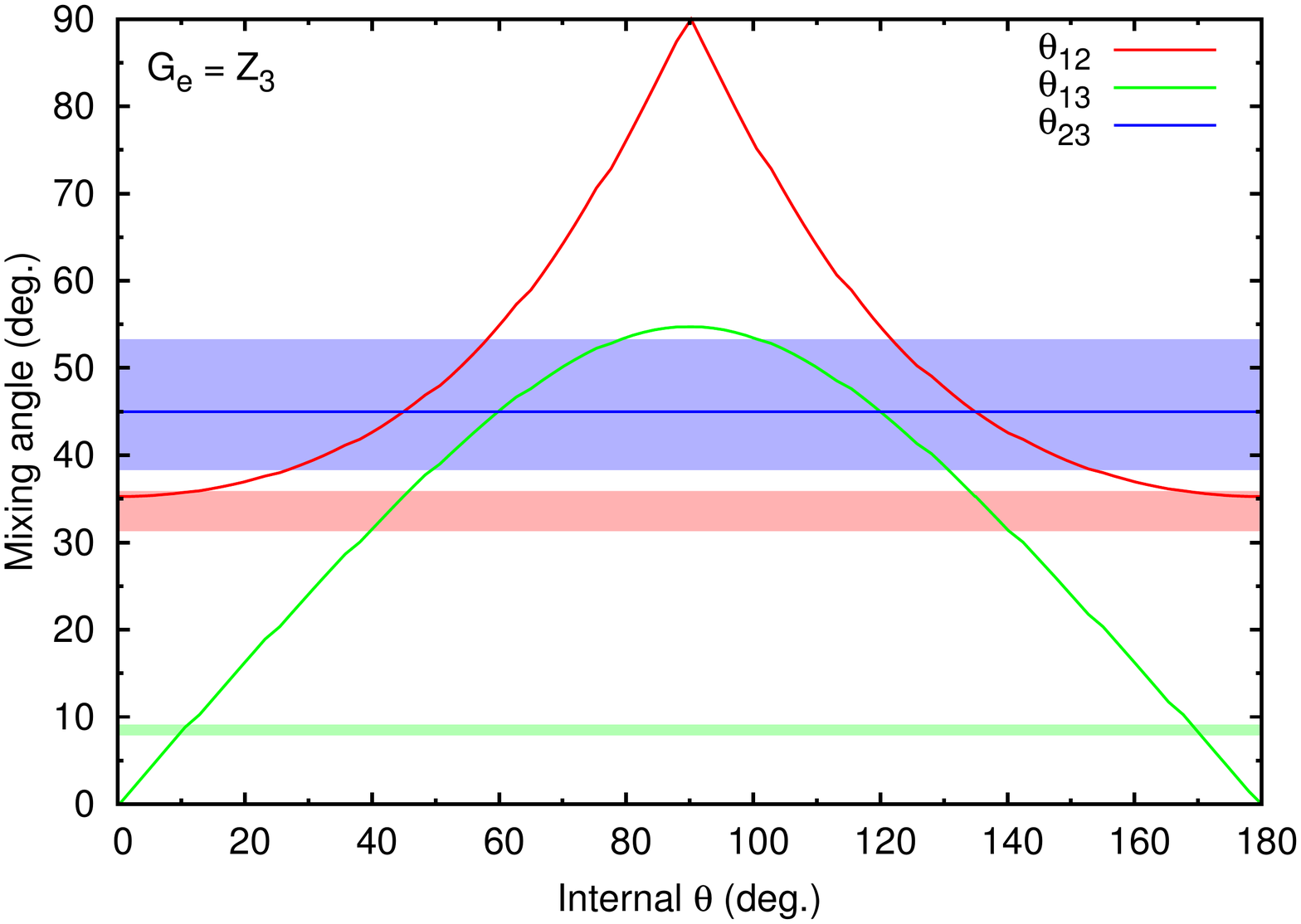}%
\includegraphics[width=0.4\textwidth]{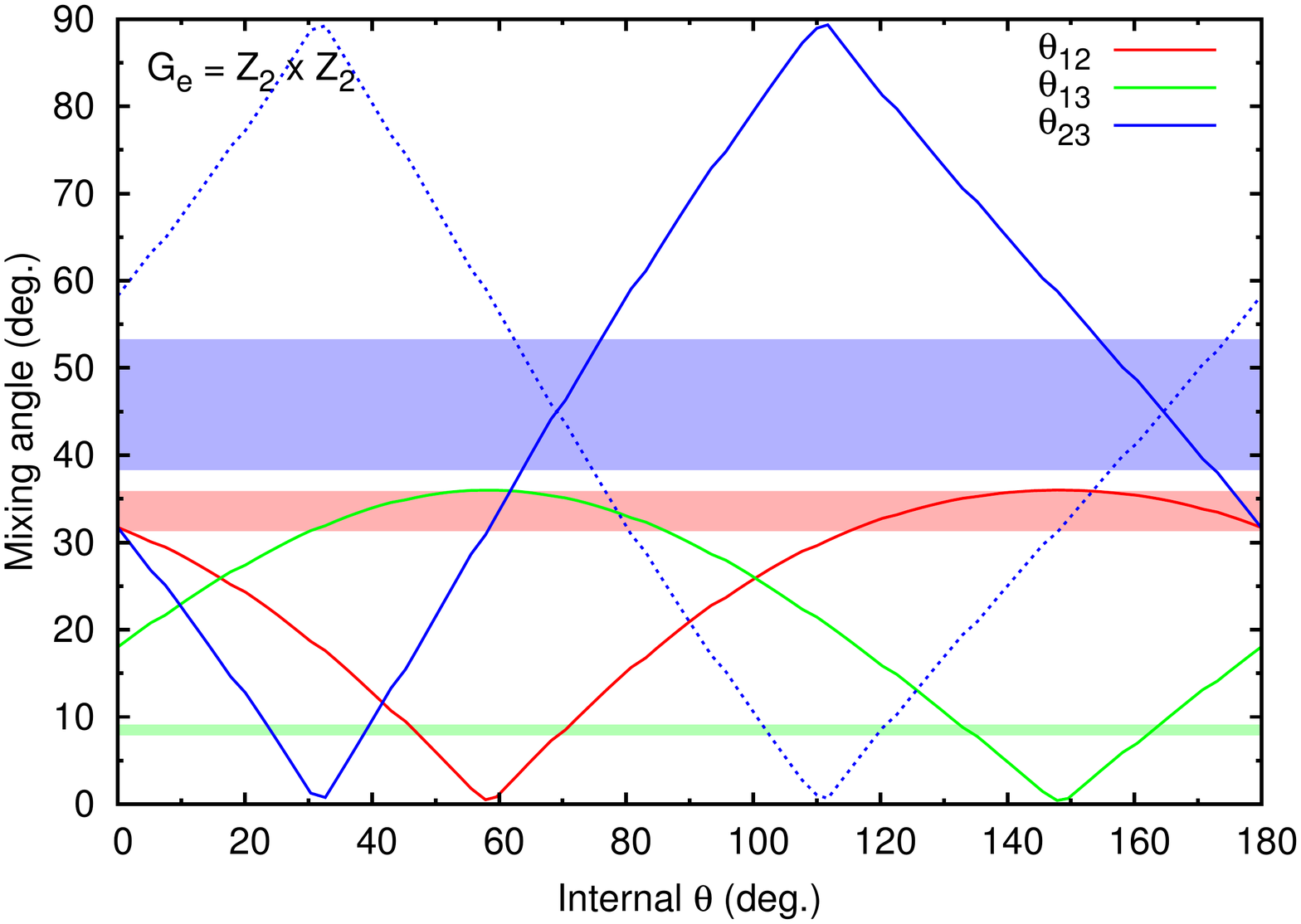}
\caption{\label{fig:Z5} Mixing patterns for $\mathbb{Z}_3$  and $\mathbb{Z}_2\times \mathbb{Z}_2$ as a function of the internal parameter $\theta$.  The shaded regions show the $3\sigma$ allowed region for the corresponding mixing angle according to current global data \cite{Gonzalez-Garcia:2014bfa}.
 }
\end{figure*}

\section{Summary}

We find that combining  $A_5$ and gCP symmetries permits a number of viable predictions of leptonic  mixing angle and phases as shown in Table~\ref{tab:predictions}. These predictions have distinct phenomenological signatures, such as maximal $\theta_{23}$ associated to maximal $\delta$, which are readily testable at the upcoming long and medium base line experiments.  

\end{document}